\documentclass[aps,prl,twocolumn,preprintnumbers,
floatfix]{revtex4}

\usepackage{amsmath,amsfonts}
\usepackage{bm}

\newcommand{\eeq}{\end{equation}}
\newcommand{\beq}{\begin{equation}}
\newcommand{\ba}{\begin{array}}
\newcommand{\ea}{\end{array}}
\newcommand{\bea}{\begin{eqnarray}}
\newcommand{\eea}{\end{eqnarray}}
\newcommand{\vev}[1]{\langle #1\rangle}

\newcommand{\eps}{\epsilon}

\newcommand{\al}{\alpha}
\newcommand{\alb}{\bar{\alpha}}

\begin{document}

\preprint{TH-UOI-04-8}

\title{Fermion masses and proton decay in string-inspired SU$(4)\times$SU$(2)^2\times$U$(1)_X$}

\author{Thomas Dent}
\author{George Leontaris}
\author{John Rizos}
\affiliation{Theoretical Physics, University of Ioannina\\ Ioannina 45110 GREECE}

\date{\today}

\begin{abstract} \noindent
We present a supersymmetric model of fermion masses with SU$(4)\times$SU$(2)^2\times$U$(1)_X$ gauge group with matter in fundamental and antisymmetric tensor representations only. The up, down, charged lepton and neutrino Yukawa matrices are distinguished by different Clebsch-Gordan coefficients due to contracting over SU$(4)$ and SU$(2)_R$ indices. We obtain a hierarchical light neutrino mass spectrum with bi-large mixing. The condition that anomalies be cancelled by a Green-Schwarz mechanism 
leads to fractional U$(1)_X$ charges which exclude $B$ violation through dimension-4 and -5 operators.
\end{abstract}


\maketitle

\noindent The values of the adjustable parameters of the Standard Model (SM) Lagrangian may be an important clue to physics beyond it (BSM). Thus, the experimentally measured values of gauge coupling constants, fermion masses and quark mixing angles, and now neutrino mass-squared differences and mixing angles (strictly, already a BSM effect), can be compared to the predictions of various types of model with full or partial \cite{Pati:1974yy} gauge unification and/or flavour symmetries.
Recent data on atmospheric and solar neutrinos \cite{Bahcall:2004ut}, implying large 1-2 and 2-3 mixing angles, present challenges for any unified framework in which neutrinos form part of a multiplet with quarks \cite{Pati:2003gv}. 

In this paper we revisit the string-inspired 4-2-2 models \cite{Antoniadis:1988cm} (see also \cite{King:1997ia}), whose implications for fermion masses were previously investigated in \cite{Allanach:1995ch,Allanach:1996hz} and which have several attractive features. Large Higgs representations (problematic to obtain in string models) are not required, the doublet-triplet splitting problem is absent, third generation fermion Yukawa couplings are unified \cite{AllanachKQuad} up to small corrections, and unification of gauge couplings is allowed and, if one assumes the model embedded in supersymmetric string, might be predicted \cite{LTracas}.

Small effective Yukawa couplings arise from nonrenormalizable superpotential operators involving a singlet charged under U$(1)_X$ \cite{Irges:1998ax} and Higgses which receive v.e.v.'s at the SU$(4)\times$SU$(2)_R$ breaking scale. A particular feature of the model is the presence of two {\em a priori}\/ independent expansion parameters depending on the sign of the U$(1)_X$ charge. For nonrenormalizable operators involving products of SU$(4)\times$SU$(2)_R$ breaking Higgses, different contractions of gauge group indices lead to effective Yukawa couplings which differ between the up ($u$) and down ($d$) quarks and the charged leptons ($e$) and neutrinos ($\nu$). This freedom is exploited to fit the different mass hierarchies in the $u$, $d$ and $e$ sectors and produce the CKM mixing. 

Right-handed neutrinos (RHNs) are automatically present and obtain Majorana masses, again through nonrenormalizable couplings to the U$(1)_X$-charged singlet and to Higgses. 
Their mass spectrum is hierarchical with the lightest RHN around $10^{11}\,$GeV and the heaviest just below the Pati-Salam breaking scale. The RHN mass hierarchy cancels off against the hierarchical neutrino Dirac mass matrix to yield a seesaw mass matrix with large off-diagonal elements. The Green-Schwarz anomaly cancellation conditions can always be satisfied, and imply fractional U$(1)_X$ charges which disallow many $B$- and $L$-violating operators.

\section{The model}
\noindent
The field content is summarized in Table 1, 
\begin{table}[here] \caption{Field content and U$(1)_X$ charges}\label{table1}
\begin{ruledtabular}
\begin{tabular}{c|cccc}
           &  SU$(4)$ & SU$(2)_L$ & SU$(2)_R$ & U$(1)_X$  \\ \hline
$F_i$      &     4    &     2     &     1     &  $\al_i $ \\
$\bar{F}_i$& $\bar{4}$&     1     & $\bar{2}$ & $\bar{\al}_i$ \\ 
$H$        &     4    &     1     &     2     &   $x$     \\
$\bar{H}$  & $\bar{4}$&     1     & $\bar{2}$ & $\bar{x}$ \\
$\phi$     &     1    &     1     &     1     &   $z$     \\
$h$        &     1    & $\bar{2}$ &     2     & $-\al_3-\bar{\al}_3$ \\
$D_1$      &     6    &     1     &     1     &  $-2x$    \\
$D_2$      &     6    &     1     &     1     & $-2\bar{x}$    
\end{tabular}
\end{ruledtabular}
\end{table}
where $i$ ranges from 1 to 3. The two $D$ fields are introduced
to give mass to colour triplet components of $H$ and $\bar{H}$ once their sneutrino-like components obtain v.\,e.\,v.'s as follows:
\[
\vev{H}= \vev{H_\nu} = M_{\rm G},\ \vev{\bar{H}}=\vev{\bar{H}_\nu} = M_{\rm G}. 
\]
In the stable SUSY vacuum the singlet $\phi$ 
obtains a v.\,e.\,v.\ to satisfy the D-flat condition including the anomalous Fayet-Iliopoulos term \cite{DineSW}.

In general a string model may have more than one singlet $\phi_i$ and more than one set of Higgses $H_i$, $\bar{H}_i$, with different U$(1)_X$ charge; the Higgses may obtain masses through $H\bar{H}\phi$ couplings. In order to break the Pati-Salam group while preserving SUSY we require that one $H$-$\bar{H}$ pair be massless at this level. This ``symmetry-breaking'' Higgs pair may be a linear combination of many $H_i$ and $\bar{H}_i$ in the charge basis, which will in general complicate the expressions for fermion masses (as will the presence of many $\phi_i$ fields). 

However, if we impose that all products $H_i\bar{H}_j$ have the same sign of U$(1)_X$ charge, and that all $\phi_i$ likewise have the same sign of charge (opposite to that of $H\bar{H}$), then the leading contributions to fermion masses from nonrenormalizable operators arise from the $H\bar{H}$ pair and $\phi$ field which have the smallest absolute value of U$(1)_X$ charge; other v.e.v.'s will enter at higher order and will be small corrections. 
Hence, and for simplicity, we restrict ourselves to a single copy of $H$, $\bar{H}$ and $\phi$. 

Standard Model fermion mass terms 
arise after electroweak symmetry-breaking from the operators 
\begin{multline} 
W_{D} = y_0^{33} F_3\bar{F}_3h + \delta \sum_{m>0} y_m^{ij} F_i\bar{F}_j h \left(\frac{\phi}{M_S}\right)^m \\ 
+ \delta \sum_{n>0} {y'}_n^{ij} F_i\bar{F}_j h \left(\frac{H\bar{H}}{M_S^2} \right)^n + \cdots 
\end{multline} 
where $M_S$ is the ``string scale'' which governs the suppression of nonrenormalizable terms in the effective theory, and only the 33 element 
is nonvanishing at renormalizable level. Nonrenormalizable terms may be suppressed by an overall factor $\delta$ of order 1. The couplings $y_n^{ij}$ and ${y'}_n^{ij}$ are nonvanishing and generically of order $1$ whenever the U$(1)_X$ charge of the corresponding operator
vanishes. Other higher-dimension operators may arise by multiplying any term by $(H\bar{H})^a\phi^b/M_S^{2a+b}$ for positive integer $a$ and $b$ such that $a(x+\bar{x})+bz=0$; however such terms are negligible unless the leading term vanishes.

Majorana masses arise from the operators
\begin{multline}
W_{M} 
= \frac{\bar{F}_i\bar{F}_j HH}{M_S} \left( \mu_0^{ij} +
\sum_{p>0} \mu_p^{ij} \left(\frac{\phi}{M_S}\right)^p \right. \\ \left. + \sum_{q>0} {\mu'}_q^{ij} \left(\frac{H\bar{H}}{M_S^2}\right)^q \right)
\end{multline}
Again, the U$(1)_X$ charges 
$\bar{\alpha}_i+\bar{\alpha}_j +2x+pz$ or $\bar{\alpha}_i+\bar{\alpha}_j+2x +q(x+\bar{x})$ must be zero if the couplings $\mu_p^{ij}$ and ${\mu'}_q^{ij}$, respectively, are not to vanish.

For nonrenormalizable Dirac mass terms involving $n$ products $H\bar{H}/M_S^2$ the gauge group indices may be contracted in different ways \cite{Allanach:1996hz} leading to Clebsch factors $C^{ij}_{n(u,d,e,\nu)}$ multiplying the effective Yukawa coupling: these are generically numbers of order $1$ and may be zero in some cases. Although the Clebsch coefficient for a particular operator $O_n$ may vanish at order $n$, the coefficient for the operator $O_{(n+a);b}$ containing $a$ additional factors $(H\bar{H})$ and $b$ factors of $\phi$ is generically nonzero.

Dirac mass terms at the unification scale are then
\begin{align} 
\frac{m^{ij}}{m_3} &= \delta_{i3}\delta_{j3} + \delta \frac{y_n^{ij}}{y_0^{33}} \left(\frac{\vev{\phi}}{M_S}\right)^m + \delta \frac{{y'}_n^{ij}}{y_0^{33}} C_n^{ij} \left(\frac{M_{\rm G}^2}{M_S^2}\right)^n \nonumber \\ 
&\simeq \delta_{i3}\delta_{j3} + \delta\left( \eps^{|\hat{m}|} + C^{ij} {\eps'}^{|\hat{n}|} \right)
\end{align}
where $m_3\equiv v_{u,d} y_0^{33}$ with $v_u$ and $v_d$ being the up-type and down-type Higgs v.e.v.'s respectively, and we define  
\[ 
\eps\equiv \left(\frac{\vev{\phi}}{M_S}\right)^{|1/z|},\qquad 
\eps'\equiv \left(\frac{M_{\rm G}^2}{M_S^2}\right)^{1/|x+\bar{x}|}.
\] 
We suppress higher-order terms involving products $\eps\eps'$.
Thus $\hat{m}= -( \alpha_i-\alpha_3 +\bar{\alpha}_j-\bar{\alpha}_3)$ and $\hat{n}= -(\alpha_i-\alpha_3 +\bar{\alpha}_j-\bar{\alpha}_3)$. The signs of $\hat{m}$ and $\hat{n}$ must be the same as $z$ and $x+\bar{x}$ respectively for the mass term to exist. 
Since the integers
$m$ and $n$ are always positive, we have $\eps^{|\hat{m}|}\equiv (\vev{\phi}/M_S)^m$.
Majorana mass terms are
\beq
M_M^{ij} =
M_R \left( \mu_p^{ij}\eps^{|\hat{p}|} + {\mu'}_{q}^{ij} {\eps'}^{|\hat{q}|} \right) 
 \simeq 
M_R ( \eps^{|\hat{p}|} + {\eps'}^{|\hat{q}|})
\eeq
where $M_R\equiv M_{\rm G}^2/M_S\equiv \eps'M_{S}$, the term with $p=q=0$ is understood, and
$\hat{p}= -(\bar{\alpha}_i +\bar{\alpha}_j +2x)$ and $\hat{q}= -(\bar{\alpha}_i +\bar{\alpha}_j +2x)$. The full neutrino mass matrix in the basis $(\nu, \nu^c)$ is of the ``seesaw'' form
and the resulting light neutrino mass matrix is 
\beq \label{seesaw}
m_\nu = - m_{D\nu} M_M^{-1} m_{D\nu}^T. 
\eeq 

\section{Parameter choices and mass matrices}
\noindent So far the discussion has been independent of the choice of U$(1)_X$ charges. The fermion mass terms 
are invariant under the family-independent shifts of U$(1)_X$ charge
\begin{align}
\alpha_i &\rightarrow \alpha_i + \zeta, & \bar{\alpha}_i &\rightarrow \bar{\alpha}_i + \bar{\zeta} \nonumber \\ 
x &\rightarrow x - \bar{\zeta} & \bar{x} &\rightarrow \bar{x} + \bar{\zeta} \label{zeta}
\end{align}
(as are the $HHD_1$ and $\bar{H}\bar{H}D_2$ couplings). Thus for the purpose of investigating fermion masses, we are free to assign 
$\al_3=\alb_3=0$ and apply $\zeta$ and $\bar{\zeta}$ shifts at the end of the calculation. The charges are then $x+\bar{x}=1$, $z=-1$, $\al_1=-4$, $\alb_1=-2$, $\al_2=-3$, $\alb_2=1$, which generate the charge matrices 
\begin{align} 
\label{chargem}
Q_X[M_D] &=
\begin{pmatrix}
-6 & -3 & -4 \\ -5 & -2 & -3 \\ -2 & 1 & 0 
\end{pmatrix} 
, \\
Q_X[M_M] &= 2x +
\begin{pmatrix}
-4 & -1 & -2 \\ -1 & 2 & 1 \\ -2 & 1 & 0
\end{pmatrix}.
\end{align}
We set $\delta=1$ and substitute $\eps$ and $\eps'$ by a single expansion parameter $\eta\simeq 0.06$ (or $\sqrt{\eta} \simeq 0.24$) via $\eps = \eta$, $\eps' = \sqrt{\eta}$. Thus the neutral gauge singlet $H\bar{H}\phi/M_S^3$ has a v.e.v.\ of $\eta^{3/2}\simeq 0.015$. 

We obtain a hierarchical common Dirac mass matrix and ($x$-dependent) Majorana RHN mass matrix in powers of $\eta$. We then have to specify Clebsch coefficients for  operators
involving one or more powers of $H\bar{H}/M_S^2$. By taking linear combinations of operators with different contractions over SU$(4)\times$SU$(2)_R$ indices one can obtain any vector in the space $(u,d,\nu,e)$, since the operators constitute a complete set over this space. Thus at first sight the model has little predictivity. 

However, in specific string models these coefficients are calculable in terms of cocycle factors \cite{cocycle}; in the absence of a specific string construction we impose that the $C^{ij}_{n}$ should be either small integers or simple rational numbers. We can take all $C^{ij}$ equal to unity apart from the following, where we quote coefficients up to a possible complex phase:
$C_d^{12}=C_d^{22}=\frac{1}{3}$, $C_u^{23}=3$, $C_u^{11}=C_u^{12}=C_u^{21}=C_u^{22}=C_u^{31}=0$,
all multiplying the leading nonvanishing entry.
The ratio $C_e^{22}/C_d^{22}=3$ is the usual Georgi-Jarlskog choice to fit the different ratios $m_s/m_b$ and $m_\mu/m_\tau$. As explained above, when $C^{ij}_n$ vanishes for the leading term, the next-to leading term is smaller by a factor 
$\eta^{3/2}$. Hence the Dirac mass matrices at the GUT scale take the following form, up to complex phases and factors of order 1:
\begin{multline} 
\frac{m_u}{m_{t0}} = \begin{pmatrix} 
\eta^{9/2} & \eta^3 & \eta^2 \\ 
                    \eta^4 & \eta^{5/2} & 3\eta^{3/2} \\ 
                                        \eta^{5/2} & \eta & 1   \end{pmatrix},\
\frac{m_d}{m_{b0}} = \\ \begin{pmatrix} 
\eta^3 & \frac{\eta^{3/2}}{3} & \eta^2 \\ 
                    \eta^{5/2} & \frac{\eta}{3} & \eta^{3/2} \\ 
                                        \eta & \eta & 1         \end{pmatrix},\
\frac{m_e}{m_{\tau 0}} = \begin{pmatrix} 
\eta & \eta^{3/2} & \eta^2 \\ 
                    \eta^{5/2} & \eta & \eta^{3/2}  \\ 
                                        \eta & \eta & 1         \end{pmatrix}
\end{multline}
where $m_{t0}$ is the top mass at the GUT scale and $m_{b0}=m_{\tau 0}=m_{t0}/\tan \beta$. For this simple choice, the resulting eigenvalues and quark mixings can be RG evolved to observable energies, where they yield acceptable fits. For example the CKM mixing angle $\theta_{12}$ is $\sqrt{\eta}\simeq 0.24$, while the angle $\theta_{13}$ is $\eta^2\simeq 0.0035$; the ratio $(m_u/m_t)_{|M_Z}$ is $\eta^{9/2}/\zeta^3 \simeq 6\times 10^{-6}$, where $\zeta\simeq 0.83$ accounts for the RG evolution of $y_{u,c,t}$ in the large $\tan\beta$ (fixed point) regime. No fine-tuned cancellations between unknown order 1 coefficients are needed. Such coefficients arise from the couplings $y_m^{ij}$, {\em etc.}, which are SU$(4)$ symmetric, and thus affect the Dirac mass matrix in the same way in each sector. This constraint may have consequences for a more detailed comparison with data.

The light neutrino mass matrix depends on $x$ through the Majorana RHN matrix:
we find two possible values, $x=1$ and $x=3/2$. \footnote{For $x<1$, $m_\nu$ no longer has a form consistent with bi-large mixing; if $x>3/2$ then the RHNs are too light and the light $\nu$ masses too large.} The seesaw mass matrix is then
\beq
m_\nu = \frac{(m_\tau\tan\beta)^2}{\eta^{2x}M_R} \begin{pmatrix} \eta & \sqrt{\eta} & \sqrt{\eta} \\ \sqrt{\eta} & 1 & 1 \\ \sqrt{\eta} & 1 & 1 \end{pmatrix}
\eeq
up to order 1 factors (different from those in the Dirac matrices), where we display leading terms in $\eta$. With a mildly fine-tuned (at the 20\% level) choice of order 1 coefficients one can obtain a neutrino mass spectrum with normal hierarchy and bi-large mixing. We did not find any charge assignments consistent with an inverted hierarchy or degenerate light neutrino spectrum.

If we take $x=1$, $\tan\beta=40$ and $M_R\equiv \eta^{1/4}M_{\rm G} 
$ to be $1.2\times 10^{16}\,$GeV, then the largest entries in $m_\nu$ are of order $0.12\,$eV. The spectrum of RHN's comprises two superheavy states of mass $(1\pm \eta/2)M_R$ and one light state of mass $\eta^4 M_R\simeq 1.5\times 10^{11}\,$GeV.

Alternatively, with $x=3/2$, $\tan \beta=45$ and $M_R$ being $\sqrt{\eta}$ times the reduced Planck mass $2.4\times 10^{18}\,$GeV, {\em i.e.}\ $M_R\simeq 6\times 10^{17}\,$GeV, the largest entries in $m_\nu$ are of order $0.05\,$eV. The RHN masses are now of order $\sqrt{\eta}M_R$, $\eta^{3/2}M_R$ and $\eta^5M_R\simeq 5\times 10^{11}\,$GeV. Thus the correct scale of light $\nu$ masses follows, with RHN mass terms derived from either the SUSY-GUT or heterotic string scale via nonrenormalizable operators. The lightest RHN masses are rather large for standard thermal leptogenesis \cite{Plumacher} if one takes into account gravitino production (given $m_{3/2}=1$-$10\,$TeV), but might be considered for nonthermal leptogenesis \cite{nontherm} or in case the gravitino is light or very heavy.

\section{Anomalies and $B$ and $L$ violation}
For gauge coupling unification at the string scale (up to threshold corrections) the non-Abelian gauge groups are required to have equal Ka{\v c}-Moody levels $k_4=k_{2L}=k_{2R}=1$. The U$(1)_X$ mixed anomalies can only be cancelled by a Green-Schwarz mechanism if anomaly coefficients obey the relation $A= \mbox{const.}\times k$, hence we require the 4-4-$1_X$, $2_L$-$2_L$-$1_X$ and $2_R$-$2_R$-$1_X$ coefficients to be equal: $A_4=A_{2L}=A_{2R}$. 
We have (up to an overall factor)
\bea
A_4 &=& 2\sum_i (\alpha_i + \bar{\alpha}_i) + 2(x+\bar{x}) + 2(-2x-2\bar{x}) \nonumber \\ 
A_{2L} &=& 4 \sum_i \alpha_i +2(-\alpha_3-\bar{\alpha}_3) \nonumber \\
A_{2R} &=& 4 \sum_i \bar{\alpha}_i +4(x+\bar{x}) +2(-\alpha_3-\bar{\alpha}_3)
\eea
from which  
we obtain the requirements
\bea
\alpha_3+\bar{\alpha}_3 -2(x+\bar{x}) &=& 0, \label{4-L-R} \\
\sum_i (\alpha_i - \bar{\alpha}_i) - (x+\bar{x}) &=& 0. \label{L-R}
\eea
The generation-independent shifts of Eq.~(\ref{zeta}) produce 
shifts in the LHS of Eqs.~(\ref{4-L-R}) and (\ref{L-R}), of $\zeta+\bar{\zeta}$ and $3(\zeta-\bar{\zeta})$ respectively. Thus, given an initial U$(1)_X$ charge assignment, it is always possible to satisfy the anomaly conditions, without altering the fermion mass matrices. 

For our chosen set of charges we have $\zeta=13/6$ and $\bar{\zeta}=-1/6$. The shifts make many operators fractionally charged: the $FFFF$ operator which would lead to $D=5$ proton decay has charge $\sum \al + 26/3$, which cannot be cancelled by any singlet combination of $H$, $\bar{H}$ and $\phi$ fields. The only surviving operators coupling matter to Higgs triplets are $FFD_1$ and $FF\bar{H}\bar{H}$, whose charges are both shifted by $2(\zeta+\bar{\zeta})=4$.
For $D=5$ proton decay one would also require appropriate mass terms for intermediate states, either $D_1D_1$ or $D_1\bar{H}\bar{H}$; but the charges of these operators are both shifted by $4\bar{\zeta}=-2/3$, hence one cannot construct the mass terms.

We must also verify that all Higgs triplet states are massive enough to evade bounds from $D=6$ operators. Down squark-like states $3_H$ and $\bar{3}_{\bar{H}}$ in $H$ and $\bar{H}$ respectively survive after breaking to the SM group. Mass terms for the $H$-$\bar{H}$-$D_1$-$D_2$ system follow from the superpotential operators
\beq
W_{\rm h}= HHD_1 + \bar{H}\bar{H}D_2 + D_1D_2 \frac{H\bar{H} H\bar{H}}{M_S^3};
\eeq
inserting v.e.v.'s we obtain up to factors of order $1$
\beq
W_{\rm h} = 
\begin{pmatrix}
\bar{3}_1 & \bar{3}_{\bar{H}} & \bar{3}_2 
\end{pmatrix}
\begin{pmatrix} 
M_{\rm G} & \mu_D & 0 \\ 0 & M_{\rm G} & 0 \\ 0 & 0 & \mu_D 
\end{pmatrix}
\begin{pmatrix} 
3_H \\ 3_2 \\ 3_1
\end{pmatrix}
\eeq
where $\mu_D = M_{\rm G}^4/M_S^3 \equiv \eta^{3/4}M_{\rm G}$ and
we decompose each sextet $D_{1,2}$ into a $\mathbf{3}$ and $\mathbf{\bar{3}}$ of SU$(3)$.
There are two mass eigenvalues of approximately $M_{\rm G}$ 
and one eigenvalue $\mu_D$ from the combination $\bar{3}_2$-$3_1$. Clearly there is no transition between $3_1$ and either $\bar{3}_1$ or $\bar{3}_H$, at any order.

Other $B$- and $L$-violating operators 
are
\beq
F_i\bar{H} h,\ \eps_{ab}F_{iA}\bar{F}_j^{Aa} F_{kB}\bar{H}^{Bb},\ \bar{F}_i\bar{F}_j\bar{F}_k\bar{H},\ F_iF_jF_kHh
\eeq
where we have indicated SU$(4)$ and SU$(2)_R$ summations in the second term.~\footnote{The conventional R-parity may be obtained from a $\mathbb{Z}_2$ symmetry acting either on $F$ and $\bar{F}$, or on $H$ and $\bar{H}$.} These give rise to superpotential terms $LH_u$; $QD^cL$ and $LE^cL$; $U^cD^cD^c$; and $QQQH_d$ respectively. Out of these, $F\bar{H} h$ is not shifted, hence is allowed (suppressed by some powers of $\eta$) if $\bar{x}$ is integer; $F\bar{F}F\bar{H}$ is shifted by $4$ and is allowed under the same condition; $\bar{F}^4$ is shifted by $-2/3$, hence is disallowed; and $F^3Hh$ is shifted by $14/3$, hence also disallowed. In a more general model with many $H$-$\bar{H}$ pairs and singlets, with a range of U$(1)_X$ charges, the analysis of anomaly coefficients, proton decay operators and R-parity violation will be different. But at least in the minimal version, all dangerous $B$-violating operators vanish automatically.

The ``mu-term'' of the MSSM originates from the 
product $hh$ 
whose U$(1)_X$ charge is $-4$: hence it receives only a mild suppression. Thus the ``mu-problem'' is not solved by the U$(1)_X$ symmetry alone.

In conclusion, we have presented a string-inspired supersymmetric SU$(4)\times$SU$(2)^2\times$U$(1)_X$ model with 5 discretely adjustable charges, 8 discretely adjusted Clebsch factors, 1 adjustable expansion parameter and 1 adjustable mass ratio ($v/M_{\rm G}$), which is consistent with gauge unification and with all known elementary fermion masses and mixings, if the spectrum of light neutrinos is hierarchical. The simplest version of the model is also free from $B$ violation through dimension-4 and -5 operators, but may allow $L$ violation. The lightest RHN mass is a few times $10^{11}\,$GeV, the lightest neutrino mass eigenstate is of order $\eta/2 \simeq 0.03$ times the heaviest,
and the neutrino mixing angle $\theta_{13}$ is of order $\sqrt{\eta}/2 \simeq 0.12$.

Other issues for further investigation include gauge unification, CP violation, supersymmetry-breaking and flavour-changing effects in both quark and lepton sectors, and cosmology including inflation, baryogenesis and dark matter.

\section*{Acknowledgements}
\par
This work was supported by the European Union under the RTN contract HPRN-CT-2000-00148.

\def\ijmp#1#2#3{{Int. Jour. Mod. Phys.}
{\bf #1},~#3~(#2)}
\def\plb#1#2#3{{Phys. Lett. B }{\bf #1},~#3~(#2)}
\def\zpc#1#2#3{{Z. Phys. C }{\bf #1},~#3~(#2)}
\def\prl#1#2#3{{Phys. Rev. Lett.}
{\bf #1},~#3~(#2)}
\def\rmp#1#2#3{{Rev. Mod. Phys.}
{\bf #1},~#3~(#2)}
\def\prep#1#2#3{{Phys. Rep. }{\bf #1},~#3~(#2)}
\def\prd#1#2#3{{Phys. Rev. D }{\bf #1},~#3~(#2)}
\def\npb#1#2#3{{Nucl. Phys. }{\bf B#1},~#3~(#2)}
\def\npps#1#2#3{{Nucl. Phys. B (Proc. Sup.)}
{\bf #1},~#3~(#2)}
\def\mpl#1#2#3{{Mod. Phys. Lett.}
{\bf #1},~#3~(#2)}
\def\arnps#1#2#3{{Annu. Rev. Nucl. Part. Sci.}
{\bf #1},~#3~(#2)}
\def\sjnp#1#2#3{{Sov. J. Nucl. Phys.}
{\bf #1},~#3~(#2)}
\def\jetp#1#2#3{{JETP Lett. }{\bf #1},~#3~(#2)}
\def\app#1#2#3{{Acta Phys. Pol.}
{\bf #1},~#3~(#2)}
\def\rnc#1#2#3{{Riv. Nuovo Cim.}
{\bf #1},~#3~(#2)}
\def\ap#1#2#3{{Ann. Phys. }{\bf #1},~#3~(#2)}
\def\ptp#1#2#3{{Prog. Theor. Phys.}
{\bf #1},~#3~(#2)}
\def\apjl#1#2#3{{Astrophys. J. Lett.}
{\bf #1},~#3~(#2)}
\def\n#1#2#3{{Nature }{\bf #1},~#3~(#2)}
\def\apj#1#2#3{{Astrophys. J.}
{\bf #1},~#3~(#2)}
\def\anj#1#2#3{{Astron. J. }{\bf #1},~#3~(#2)}
\def\apjs#1#2#3{{Astrophys. J., Suppl. Ser.}
{\bf #1},~#3~(#2)}
\def\mnras#1#2#3{{MNRAS }{\bf #1},~#3~(#2)}
\def\grg#1#2#3{{Gen. Rel. Grav.}
{\bf #1},~#3~(#2)}
\def\s#1#2#3{{Science }{\bf #1},~#3~(#2)}
\def\baas#1#2#3{{Bull. Am. Astron. Soc.}
{\bf #1},~#3~(#2)}
\def\ibid#1#2#3{{\it ibid. }{\bf #1},~#3~(#2)}
\def\cpc#1#2#3{{Comput. Phys. Commun.}
{\bf #1},~#3~(#2)}
\def\astp#1#2#3{{Astropart. Phys.}
{\bf #1},~#3~(#2)}
\def\epjc#1#2#3{{Eur. Phys. J. C}
{\bf #1},~#3~(#2)}
\def\nima#1#2#3{{Nucl. Instrum. Meth. A}
{\bf #1},~#3~(#2)}
\def\jhep#1#2#3{{J. High Energy Phys.}
{\bf #1},~#3~(#2)}
\def\lnp#1#2#3{{Lect. Notes Phys.}
{\bf #1},~#3~(#2)}
\def\appb#1#2#3{{Acta Phys. Pol. B}
{\bf #1},~#3~(#2)}
\def\yf#1#2#3{{Yad. Fiz.}
{\bf #1},~#3~(#2)}


\begin{thebibliography}{99}



\bibitem{Pati:1974yy}
J.\,C.~Pati and A.~Salam,
Phys.\ Rev.\ D {\bf 10}, 275 (1974).

\bibitem{Bahcall:2004ut}
For recent reviews, see
J.\,N.~Bahcall, M.\,C.~Gonzalez-Garcia and C.~Pe{\~ n}a-Garay,
arXiv:hep-ph/0406294;
G.~Altarelli,
arXiv:hep-ph/0405182.

\bibitem{Pati:2003gv}
J.\,C.~Pati,
Phys.\ Rev.\ D {\bf 68}, 072002 (2003).

\bibitem{Antoniadis:1988cm}
I.~Antoniadis and G.\,K.~Leontaris,
Phys.\ Lett.\ B {\bf 216}, 333 (1989);
I.~Antoniadis, G.\,K.~Leontaris and J.~Rizos,
Phys.\ Lett.\ B {\bf 245}, 161 (1990);
G.\,K.~Leontaris and J.~Rizos,
Nucl.\ Phys.\ B {\bf 554}, 3 (1999).

\bibitem{King:1997ia}
S.\,F.~King and Q.~Shafi,
Phys.\ Lett.\ B {\bf 422}, 135 (1998);
R.~Jeannerot, S.~Khalil, G.~Lazarides and Q.~Shafi,
JHEP {\bf 0010}, 012 (2000);
S.\,F.~King and G.~G.~Ross,
Phys.\ Lett.\ B {\bf 574}, 239 (2003).

\bibitem{Allanach:1995ch}
B.\,C.~Allanach and S.\,F.~King,
Nucl.\ Phys.\ B {\bf 459}, 75 (1996).

\bibitem{Allanach:1996hz}
B.\,C.~Allanach, S.\,F.~King, G.\,K.~Leontaris and S.~Lola,
Phys.\ Rev.\ D {\bf 56}, 2632 (1997).

\bibitem{AllanachKQuad}
B.\,C.~Allanach and S.\,F.~King,
Phys.\ Lett.\ B {\bf 353} 477 (1995).

\bibitem{LTracas}
G.\,K.~Leontaris and N.\,D.~Tracas,
Phys.\ Lett.\ B {\bf 372} 219 (1996).


\bibitem{Irges:1998ax}
N.~Irges, S.~Lavignac and P.~Ramond,
Phys.\ Rev.\ D {\bf 58}, 035003 (1998);
G.~Altarelli and F.~Feruglio,
Phys.\ Lett.\ B {\bf 439}, 112 (1998);
J.\,R.~Ellis, G.\,K.~Leontaris, S.~Lola and D.\,V.~Nanopoulos,
Eur.\ Phys.\ J.\ C {\bf 9}, 389 (1999);
G.\,K.~Leontaris and J.~Rizos,
Nucl.\ Phys.\ B {\bf 567}, 32 (2000).

\bibitem{DineSW}
M.~Dine, N.~Seiberg and E.~Witten,
Nucl.\ Phys.\ B {\bf 289} 589 (1987).

\bibitem{cocycle}
V.\,A.~Kostelecky, O.~Lechtenfeld, W.~Lerche, S.~Samuel and S.~Watamura,
Nucl.\ Phys.\ B {\bf 288} 173 (1987).

\bibitem{Plumacher} See {\em e.g.}\
W.~Buchm{\" u}ller, P.~Di Bari and M.~Pl{\" u}macher,
arXiv:hep-ph/0406014.

\bibitem{nontherm}
G.~Lazarides and Q.~Shafi,
Phys.\ Lett.\ B {\bf 258}, 305 (1991);
T.~Dent, G.~Lazarides and R.~Ruiz de Austri,
Phys.\ Rev.\ D {\bf 69} 075012 (2004).

\end{thebibliography}
\end{document}